\documentclass{emulateapj}
\slugcomment{{\sc Accepted to ApJ:} January 4, 2008}

\def\arcsec{{$^{\prime\prime}$}}
\def\ptsec{$^{\prime\prime}\mskip-7.6mu.\,$}

\shorttitle{Quiescent H$_2$ Emission}
\shortauthors{Bary et al.}

\begin{document}

\newcommand\tna{\,\tablenotemark{a}}
\newcommand\tnb{\,\tablenotemark{b}}
\newcommand\tnc{\,\tablenotemark{c}}
\newcommand\tnd{\,\tablenotemark{d}}
\newcommand\tne{\,\tablenotemark{e}}
\newcommand\tnf{\,\tablenotemark{f}}

\title{Quiescent H$_2$ Emission From Pre-Main Sequence Stars in Chamaeleon I}

\author{Jeffrey S. Bary$^1$, David A. Weintraub$^2$, Sonali J. Shukla$^2$, Jarron M. Leisenring$^1$}
\author{Joel H. Kastner$^3$} 

\affil{$^1$Department of Astronomy, University of Virginia, P.O. Box 400325, 
Charlottesville, VA 22904-4325}
\affil{$^2$Department of Physics \& Astronomy, Vanderbilt University, P.O. Box 1807 Station B,
Nashville, TN 37235}
\affil{$^3$Carlson Center for Imaging Science, RIT, 54 Lomb Memorial Drive,
Rochester, NY 14623}
\email{jsb3r@virginia.edu,\\
david.a.weintraub@vanderbilt.edu,sonali.j.shukla@vanderbilt.edu,\\
jarron@virginia.edu,jhkpci@cis.rit.edu}

\begin{abstract}

We report the discovery of quiescent emission from molecular hydrogen gas located in the circumstellar disks of six pre-main sequence stars, including two weak-line TTS, and one Herbig AeBe star, in the Chamaeleon~I star forming region.  For two of these stars, we also place upper limits on the $2\rightarrow1~S(1)$/$1\rightarrow0~S(1)$ line ratios of $\sim$~0.4 and 0.5. Of the 11 pre-main sequence sources now known to be sources of quiescent near-infrared hydrogen emission, four possess transitional disks, which suggests that detectable levels of H$_2$ emission and the presence of inner disk holes are correlated.  These H$_2$ detections demonstrate that these inner holes are not completely devoid of gas, in agreement with the presence of observable accretion signatures for all four of these stars and the recent detections of [Ne~{\scshape ii}] emission from three of them.  The overlap in [Ne~{\scshape ii}] and H$_2$ detections hints at a possible correlation between these two features and suggests a shared excitation mechanism of high energy photons.  Our models, combined with the kinematic information from the H$_2$ lines, locate the bulk of the emitting gas at a few tens of AU from the stars.  We also find a correlation between H$_2$ detections and those targets which possess the largest H$\alpha$~equivalent widths, suggesting a link between accretion activity and quiescent H$_2$ emission.  We conclude that quiescent H$_2$ emission from relatively hot gas within the disks of TTS is most likely related to on-going accretion activity, the production of UV photons and/or X-rays, and the evolutionary status of the dust grain populations in the inner disks.

\end{abstract}


\section{Introduction}
\label{intro}

The reservoirs of gas and dust found orbiting most sun-like stars with ages on the order of a million years --- also known as T Tauri stars (TTS) --- are now subjects of intense study, as they likely hold the secrets to the formation history of the solar system and other planetary systems.  The evolution of the observable characteristics of these planet forming disks and the underlying phenomena responsible for their detectability may provide the necessary insight to further our understanding of planet forming processes and to determine the likelihood of the formation of planetary systems around sun-like stars.

The existence of circumstellar disks accompanying young low mass stars previously has been inferred by the presence of detectable emission from trace constituents of these disks (i.e., dust and gaseous molecules other than H$_2$).  The earliest attempts to determine the fraction of stars that possess disks in young star forming clusters focused on their near-infrared continuum \citep{stro1989} and submillimeter continuum emission \citep{wein1989,beck1990}.  Circumstellar disks possessing warm ($T$~$\sim$~1000~K) micron-sized dust grains within $\sim$~1~AU of the pre-main sequence stars produce near-infrared emission in excess of that predicted for the stellar photosphere, while circumstellar disks possessing cold, micron-sized dust grains at radii of several tens of AU produce excess radiation at mid-infrared and submillimeter wavelengths.  From surveys such as these, astronomers have determined the lifetime of warm dust grains in the inner disks to be 3-6~Myr \citep{lada2000,hais2000,hais2001a}.  Similarly, observations of cold ($T$~$\sim$~10-100~K) gas molecules (e.g., CO, HCN$^+$, etc) from the outer regions of the disks \citep{beck1993,zuck1995} tend to support the disk lifetime estimates drawn from near-infrared surveys.

More recently, new disk fraction and lifetime surveys of TTS in nearby star forming regions have been conducted utilizing the Infrared Array Camera (IRAC) and Multiband Imaging Photometer (MIPS) aboard the {\it Spitzer Space Telescope}.  These instruments are capable of detecting emission from dust grains ($T$~$\sim$~100-300~K) located at intermediate disk radii, a region of these disks to which the previous photometric surveys were insensitive.  \cite{ciez2007} took a census of 230 sources identified as weak-line TTS -- weakly or non-accreting TTS possessing either a passive, optically thin disk or no circumstellar material -- in the Lupus, Ophiuchus, and Perseus molecular clouds.  They find a higher disk frequency for the sources located in the highest extinction regions of the clouds, suggesting that the wTTS at the edges of the formation regions are slightly older and more evolved.  Although they find $\sim$~20\% of these possess some circumstellar material, $\sim$~50\% of what appear to be the youngest stars in the sample show no evidence of disks.  These results, which are consistent with other {\it Spitzer} surveys of low mass star forming regions \citep{furl2006,geer2006,kess2006,padg2006,damj2007}, lead them to conclude, based on the existence of detectable emission from cool dust grains, that disk lifetimes are less than ten million years, which roughly agrees with the conclusions about disk lifetimes drawn from near-infrared surveys.  Given the chaotic environment near these young stars and the phenomena (i.e., accretion, photoevaporation, stellar and disk winds, close encounters with other cluster members) that conspire to destroy these disks, the lifetimes measured from these surveys can be explained by disk dissipation models including all or some of these processes \citep{holl2000}.

Despite the growing consensus that disk lifetimes are only a few million years that has been drawn from several different techniques for studying circumstellar disks, astronomers must remain cautious when inferring the presence or absence of disks based upon the detectability of disk tracers (i.e., dust and CO emission) because the majority of the disk mass is composed of molecular hydrogen (H$_2$).  As these disks evolve according to the core accretion model of planet formation, the small micron-sized dust grains accumulate into larger bodies, which will change the wavelength dependence of the dust emissivities and the collective surface area of dust grains emitting at wavelengths in the near-infrared.  Since a planetesimal must reach a mass of $\simeq$~10 earth masses before it can gravitationally collect H$_2$ gas from the disk, the main component of the disk must persist, while planetesimals accrete into km-sized objects and disk tracers disappear, in order for gas giant planets to be the result of the core accretion scenario of planet formation.  Therefore, in an accretionally-evolved system in which small dust grains have coagulated into larger planetesimals, an evolutionary stage should exist during which both the near-infrared and mid-infrared excesses will have diminished substantially while the massive gaseous component remains present.

In previous work, \cite{wein2000}, \cite{bary2002}, and \cite{bary2003a} searched a variety of TTS, predominantly wTTS, in the nearby Taurus-Auriga, $\rho$ Ophiuchus, and TW Hya Association star forming regions for emission signatures from H$_2$ persisting in such accretionally evolved circumstellar disks.  Using various high-resolution, near-infrared spectrometers, narrow line emission centered at the $v=1\rightarrow0~S(1)$ transition rest wavelength of 2.1218~$\mu$m was detected from several TTS stars; some were known to possess circumstellar disks (TW~Hya, GG~TauA, and LkCa~15), and one was previously thought to be devoid of any circumstellar material (DoAr~21).  The detection of quiescent emission from the disk of a wTTS with no other detectable disk tracers suggests that the gaseous component may indeed persist as the disk evolves.  However, we estimate the mass of hot H$_2$ gas producing the observed features in these sources to be in the range 10$^{-12}$-10$^{-10}$~M$_\odot$.  For at least three of these four sources, a substantial reservoir of H$_2$ gas exists in their disks suggesting that the near-infrared H$_2$ emission feature, is itself only a tracer of the disk gas.  In the time since these detections were made, several other authors searching for the same emission signature also have detected H$_2$ gas \citep{itoh2003,taka2004,rams2007}.  In two of the three additional sources toward which H$_2$ was detected (LkH$\alpha$~264 and ECHA J0843.3-7905), the emission lines were similarly centered at 2.1218~$\mu$m, with full-width at half of maximum (FWHM) only slightly larger than those reported for the previous four detections.  In the case of DG~Tau, \cite{taka2004} reported the detection of an H$_2$ emission line blueshifted by 15~km~s$^{-1}$ and possessing a FWHM of $\sim$~35~km~s$^{-1}$.  The authors also show that the emitting gas is extended along the axis of a strong outflow previously observed from this source.  Therefore,  the emitting gas from this source is most likely associated with the outflowing gas distinguishing this detection from the others.  In addition to these detections of rovibrational emission, \cite{sher2003}, \cite{bitn2007} and \cite{mart2007} have detected H$_2$ emission from the mid-infrared pure rotational transitions from the Herbig AeBe stars, AB~Aur and HD~97048, with a similarly quiescent kinematic signature.

Here, we present five more detections of H$_2$ near-infrared emission from both classical and weak-line TTS, an infrared companion (IRC), and a detection from a Herbig AeBe star in the Chamaeleon I star forming region.  In addition, we place an upper limit on emission from a second line of H$_2$ at 2.2477~$\mu$m for two of these sources.  We review the possible stimulation mechanisms with respect to recent X-ray and UV irradiated disk models.  We find that two of our past detections and one of the present detections possess transitional disks --- disks with large inner disk holes devoid of emission from small warm dust grains.  Two of these transitional disk sources also possess detectable emission from a mid-infrared [Ne~{\scshape ii}] emission line predicted to be a tracer of high energy photons and a signature of photoevaporation, a mechanism thought to be potentially responsible for the presence of the inner disk holes of the transitional disks.  Kinematic information from the H$_2$ emission features suggest that the emitting gas extends to within a few AU of each source, demonstrating for the sources with inner disk holes that gas remains despite the lack of dust emission.  We estimate the location of the regions within the disks where potential excitation photons deposit their energies and find agreement with the location of emitting gas determined from the velocity dispersion of the line.  Finally, we relate the detection of quiescent H$_2$ emission to accretion activity and, possibly, the evolutionary status of the dust grain population.

\section{Observations}

We obtained high spectral resolution (R~$\simeq$~60,000) near-infrared spectra of thirteen TTS (Table~1) in the Chamaeleon~I star forming region on 2003 January 7-18, UT, using the Phoenix spectrometer \citep{hink1998} in queue-mode on the Gemini South 8-m telescope atop Cerro Pachon in Chile.  Phoenix utilizes a 256~$\times$~1024 pixel section of a 1024~$\times$~1024 pixel InSb Aladdin II detector.  Most of our spectral observations were centered at 2.1218~$\mu$m and were made using a 14\arcsec\ long, 0\ptsec17 (2 pixel) wide slit, which result in nominal instrumental spatial and velocity resolutions of 0\ptsec05 and $\sim$~5 km s$^{-1}$, respectively.  For the observations of the two faintest sources, Sz~33 and CHX~20E, and follow-up observations centered at 2.2477~$\mu$m of CS~Cha and HD~97048, we used the larger, 0\ptsec34 (4 pixel) wide slit, resulting in a velocity resolution of $\sim$~6~km~s$^{-1}$.  Spectra centered at 2.1218 $\mu$m provided a spectral coverage from 2.1170 to 2.1262~$\mu$m, while those centered at 2.2477~$\mu$m covered 2.2426 to 2.2534~$\mu$m.  The spectral region centered at the wavelength for the $v~=~1\rightarrow0~S(1)$ H$_2$ transition includes three bright telluric OH lines, at 2.11766, 2.12325, and 2.12497~$\mu$m,  while the spectra centered at the wavelength for the $v=2\rightarrow1~S(1)$ transition contains two OH lines at 2.24599 and 2.25192~$\mu$m.  These OH lines were used to provide an absolute wavelength calibration for each observation.

\begin{deluxetable}{lccccc}
\tablewidth{0pt}
\tablecaption{Cha I Source List}
\tablehead{
			\colhead{Source}
		&       \colhead{Class\tna}	
		&	\colhead{Spectral}
		&	\colhead{H$\alpha$ EW\tnc}
		&	\colhead{Log L$_{\rm x}$\tnd}
		&	\colhead{A$_{\rm v}$\tne} \\
		         \colhead{}
		&      \colhead{}        
		&	\colhead{Type\tnb} 
		&	\colhead{(\AA)}
		&	\colhead{(ergs s$^{-1}$)}
		& 	\colhead{}  \\
}

\startdata

CS Cha        			& cTTS                 & K4           & -13.3, -20, -65       &   30.0     &  0.85    \\
HD 97408    			& AeBe                 & B9.5        & -30.0                    &   29.0     &  1.30      \\
Sz 33	    			& cTTS                 & M0           & -9.5         	       &   28.9      &   1.45    \\
Glass Ia        			& wTTS                & K4           & -4.0             	       &   30.1      &  1.92    \\
Glass Ib        			& IRC                    & G5          & -9.3       	       & \nodata   &   \nodata  \\
CV Cha		                  & cTTS                  & G8          & -60.5      	       &   30.1     &  1.67    \\
Sz 41            			& wTTS                & K0-K1    &  -2.7       	       &   30.2      &   1.17    \\
LH$\alpha$ 332-17		& cTTS                 & G2          & -17.2     	       &   29.9     &    2.35  \\
CHX 20E				& wTTS                & K4.5       & 0.0      	                &   29.4    &    0.31   \\
Sz 30				& wTTS                & M0          & -5.2  	     	      &    30.1     &   1.18    \\ 
HD 97300				& AeBe                 & B9          &   0           	      &    29.2       &   1.80      \\
Ced110 IRS 2			& wTTS                & G2          &  $>$-1.2              &    30.7       &    3.39   \\
WW Cha				& cTTS                 & K5          & -67.4       	      &    29.8      &    3.20  \\

\enddata
\tablenotetext{a}{Class of stars taken from \cite{feig1993}, \cite{hama1992b}, and \cite{natt2000}.}
\tablenotetext{b}{Spectral types taken from \cite{feig1989}, \cite{gauv1992}, \cite{hama1992a}, and \cite{hama1992b}.}
\tablenotetext{c}{H$\alpha$ EWs from \cite{appe1983},  \cite{feig1989}, \cite{gauv1992}, \cite{walt1992}, and \cite{luhm2004}.}
\tablenotetext{d}{${\rm Log}$ L$_{\rm x}$ values taken from \cite{feig1993}.}
\tablenotetext{e}{Visual extinction values taken from \cite{gauv1992}.}

\end{deluxetable}

Using integration times of 900~s for our program stars (Table~1), we obtained signal-to-noise ratios (SNR) in the continua for all sources of $\ge$~50, with the exception of the continuum for Sz~30, for which the SNR was lower.  An integration time of 900 s also was used for the two observations of CS~Cha and HD~97048, centered on the 2.2477~$\mu$m H$_2$ line.  Our observing program included repeated 120~s observations of telluric calibration sources -- HR~3685 (A2IV), HR~5571 (B2III), HR~4798 (B2V), HR~5231 (B2.5IV), and HR~1956 (B7IV) -- possessing few, if any, photospheric features in this $K-$band portion of their spectra.  Similar observations of HR~5231 and HR~5762 (A2IV) centered at 2.2477~$\mu$m used as telluric calibrators were made with shorter integrations times of 60~s and 40~s, respectively.  Flat-field images were obtained using the Gemini Facility Calibration Unit Quartz Halogen lamp \citep{rams2000}.  Observations were made by nodding the telescope 5\ptsec0 along the slit, producing image pairs that were then subtracted to remove sky background and dark current.  The spectra were extracted from 0\ptsec25 north to 0\ptsec25 south of the central row of pixels of the spectral beam, corresponding to the FWHM of the spectral beam.  The target star spectra were then ratioed with one of the telluric calibration sources in order to remove telluric absorption features.

\section{Results}
\label{res}

We have detected quiescent 2.1218~$\mu$m line emission from H$_2$ from four additional TTS (CS~Cha, Sz~33, CV~Cha, and Glass Ia), one IRC (Glass~Ib, the secondary component to one of the TTS detections), and one Herbig AeBe star (HD~97048) in the Chamaeleon~I star forming region.  We define quiescent H$_2$ emission as emission at low velocities relative to and centered on the stellar rest velocity.  These detections nearly double the total number of TTS known to have quiescent H$_2$ emission at this wavelength.  Full spectra of both the detections and non-detections obtained with Phoenix are presented in Figure~1.  For each source with detectable emission, with the exception of the binary system, Glass I, the emitting gas is determined to be within a single Gaussian-shaped spectral beam with a full-width of $\sim$~1\arcsec\, placing the emission within $\sim$~80~AU of each of the sources (Cha~I; D~$\sim$~160~pc) .  In the spectral images of Glass~Ia and Ib, and the subsequent spectral extractions of Glass~I, we find that the emission from this source extends between the components of the binary, to the south of the primary, but not north of the secondary (see \S\ref{glassI}).

All spectra have been shifted to the local standard of rest (LSR) velocity frame.  Assuming a V$_{\rm LSR}$~$\sim$~2~km~s$^{-1}$ for the Cha~I sources \citep{jame2006}, we find that in all cases, the detected emission lines are centered at the rest wavelength of the $v=1\rightarrow0~S(1)$ transition within the estimated measurement error ($\pm$~1~km~s$^{-1}$).  In most cases, we find that the telluric correction was very effective, removing a significant number of telluric features.  The remaining absorption lines are expected to be associated with the stellar photospheres of the target stars.  As was found in previous detections of H$_2$ by \cite{bary2002} and \cite{bary2003a}, the H$_2$ features are resolved at this resolution.  The FWHM for these lines fall within the range of $\sim$~12~$\le$~v~$\le$~18~km~s$^{-1}$ for the five TTS and slightly higher, $\sim$~20.4~km~s$^{-1}$ for the Herbig AeBe star.  The H$_2$ equivalent widths and extinction corrected line fluxes are presented in Table~2 for the detections and one `possible' detection (WW~Cha whose spectrum is presented with the non-detections in Figure~1), along with the 3$\sigma$ upper limits for the five sources that were clearly non-detections.  The line fluxes were dereddened using A$_{\rm v}$ values from \cite{gauv1992} and a standard A$_\lambda$~$\sim$~$\lambda^{-1.84}$ extinction law adopted from \cite{mart1990}.

\begin{deluxetable}{lcccc}
\tablewidth{0pt}
\tablecaption{Cha I H$_2$ Detections and Non-Detections}
\tablehead{
			\colhead{Source}
		&	\colhead{K$_{\rm mag}$\tna}
		&	\colhead{Line Flux\tnb}
		&	\colhead{Eq. Width}
		&	\colhead{FWHM} \\
			\colhead{} 
		&	\colhead{}
		&	\colhead{(ergs cm$^{-2}$ s$^{-1}$)}
		& 	\colhead{(\AA)}
		&	\colhead{(km s$^{-1}$)} \\
}

\startdata
Detections\\
\hline
CS Cha        			& $8.26$  & 2.1$\pm$0.1$\times$10$^{-15}$         &   0.10    &  12.6    \\
HD 97408    			& $6.27$  & 8.6$\pm$0.4$\times$10$^{-15}$         &   0.07    &  20.4    \\
Sz 33	    			& $9.24$  & 6.9$\pm$0.3$\times$10$^{-16}$         &   0.08    &  12.7    \\
Glass Ia        			& $8.58$  & 2.3$\pm$0.1$\times$10$^{-15}$         &   0.43    &  12.1    \\
Glass Ib        			& $7.35$  & 6.7$\pm$0.3$\times$10$^{-15}$         &   0.05    &  17.6    \\
CV Cha                                & $6.79$  & 2.0$\pm$0.2$\times$10$^{-15}$         &    0.02   &  12.6    \\
\hline
Possible Detection\\
\hline
WW Cha				& $6.13$  & $\sim$3.7$\times$10$^{-15}$ & 0.02  &  11.8   \\
\hline
Non-Detections\\
\hline
Sz 41            			& $7.88$  & $<$1.3$\times$10$^{-15}$  &   \nodata         &    \nodata       \\
LH$\alpha$ 332-17		& $6.26$  & $<$3.0$\times$10$^{-15}$  &     \nodata       &     \nodata      \\
CHX 20E				& $9.19$  & $<$3.9$\times$10$^{-16}$  &     \nodata       &    \nodata       \\
Sz 30				& $8.55$  & $<$3.1$\times$10$^{-15}$  &     \nodata       &   \nodata        \\ 
HD97300				& $7.19$  & $<$1.9$\times$10$^{-15}$  &   \nodata         &   \nodata        \\
Ced110 IRS 2			& $6.43$  & $<$6.8$\times$10$^{-15}$  &   \nodata         &   \nodata        \\

\enddata

\tablenotetext{a}{$K$-band magnitudes taken from \cite{laws1996} with the exception of Glass Ia and Ib found in \cite{chel1988}.}
\tablenotetext{b}{The uncertainties in the line fluxes were determined based upon the continuum noise levels and the integrated line strengths.}

\end{deluxetable}

\begin{figure}[htbp]
\begin{center}
\includegraphics[width=1.0\columnwidth]{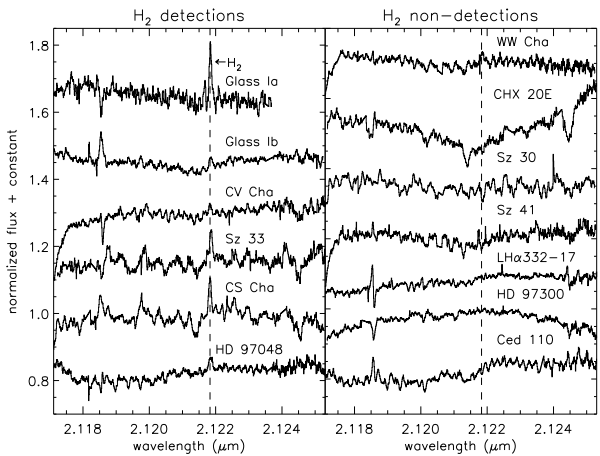}
\caption{In the left frame, we plot the spectra for sources containing H$_2$ emission line features.  The spectra have been telluric corrected and shifted into the V$_{\rm LSR}$ reference frame.  We find that spectral features center on the rest wavelength of 2.1218~$\mu$m for the $v=1\rightarrow0~S(1)$ transition of H$_2$ in the stellar rest frame, to within errors for the accepted V$_{\rm LSR}$ of Cha~I members.  In the right frame, we plot the spectra of the non-detections.  These spectra have also been telluric corrected and shifted in the V$_{\rm LSR}$ reference frame of the sources.  We note the inclusion of the spectra of WW~Cha with the non-detections, but list it as a `possible' detection in Table~2 due to the presence of a 2.4$\sigma$ feature similarly centered at 2.1218~$\mu$m.}
\label{spectra}
\end{center}
\end{figure}

In observations centered on the $v=2\rightarrow1~S(1)$ rovibrational transition of H$_2$ at 2.2477~$\mu$m of two 2.1218~$\mu$m line emission sources, CS~Cha and HD~97048, we did not detect line emission.  The upper limits to the $v=2\rightarrow1~S(1)$ line fluxes for CS~Cha and HD~97048 are 6.5~$\times$~10$^{-16}$ and 4.4~$\times$~10$^{-15}$ ergs~cm$^{-2}$~s$^{-1}$, respectively.

\section{Discussion}
\label{disc}

The lack of bulk motion of emitting H$_2$ gas along the line of sight and with respect to the systemic velocity of the stars is interpreted as evidence that the gas is not entrained in a fast moving wind or outflow associated with any of these young sources.  Additionally, the lack of extended emission beyond the full-width of the spectral beam indicates that the emitting gas is confined to regions within $\sim$~80~AU of the stars --- with the exception of binary Glass Ia and Ib, in which the emission extends between the companions.   Therefore, we conclude that, in all cases except Glass~I, the emitting H$_2$ is quiescent gas residing in the circumstellar disks associated with these sources, and is not shocked, spatially extended gas, as has been observed toward other Class~I sources and TTS (e.g., T Tauri, DG~Tau; \cite{herb1996}, \cite{taka2004}, \cite{beck2007}).  The Glass~I binary appears to be an exceptional source with its extended emission.

The small, but resolved FWHM measured for the quiescent emission features must be broadened by the Keplerian rotation of the gas as it orbits the central sources, since thermal broadening would require temperatures $>$~6000~K.  Using the same technique \cite{carr2001} and \cite{naji2003} used to estimate the inner truncation radius of the gas disk using CO emission features, we used the velocity shift of H$_2$ gas at the outermost edges of the Gaussian-shaped features, referred to as the half-width at zero intenisty (HWZI), to estimate the innermost radius at which the detected H$_2$ gas resides.  Assuming an inclination angle of 45$^\circ$, the main sequence mass corresponding to the spectral type of each source, and estimating the maximum orbital velocity of gas from the HWZI, we calculated the inner radii for gas in the disks of CS~Cha and Sz~33 to be 2.5 and 0.9~AU, respectively, while for the more massive stars, CV~Cha and HD~97048, we find 3.2 and 5.5~AU.  We note that these radii represent the innermost regions of the disks from which the emission arises and accounts for only a fraction of the overall line emission.  We conclude that the bulk of the emitting gas resides outside of these disk radii.

The results we have found for the distances from the stars at which the hot H$_2$ emission is generated is similar to that found by \cite{bary2003a}, who found FWHM of 9-14~km~s$^{-1}$ for several K type stars, placing the bulk of the emitting gas at about 10~AU, by \cite{rams2007}, who found a line width of $\sim$~18~km~s$^{-1}$ for a single M3 star (ECHAJ0843.3-7905) and by \cite{bitn2007}, who found a line width of 8.5~km~s$^{-1}$ for a single AeBe star (AB~Aur).  In all of these cases, the H$_2$ gas emitting in rovibrational or pure rotational lines is not found close to the star (inside of 1~AU) nor at great distances from the stars, suggesting that processes exciting this emission is most effective at intermediate distances of a few to a few tens of AU from these stars.

Assuming that the emitting H$_2$ gas is optically thin, we have used Equation~2 from \cite{bary2003a}, the measured line fluxes (see Table~2), and a distance of 160~pc to Chamaeleon~I, to calculate the masses of `hot' H$_2$ gas producing the line emission.   We find that these masses fall within the range 10$^{-11}-10^{-10}~{\rm M}_\odot$, in agreement with previous estimates \citep{bary2003a,rams2007}.  On their own, these masses do little for informing us about the total mass of gas in the system or the total disk mass unless we understand the physical conditions of the gas and/or the mechanism responsible for stimulating the emission.  We will address this problem in both \S\ref{ratio} and \S\ref{emr}.

\subsection{$v = 2\rightarrow1~S(1)/v = 1\rightarrow0~S(1)$}
\label{ratio}

The $v=2\rightarrow1~S(1)/v=1\rightarrow0~S(1)$ ratio has often been used to distinguish between two commonly observed mechanisms for stimulating H$_2$ emission: shock heating and UV fluorescence.  The value of this ratio for a parcel of shock heated H$_2$ gas has been observed to be $\sim$~0.1, while a value of $\sim$~0.5 has been observed for UV fluorescent H$_2$ emission from low density gas ($n_{\rm H}$~$\le$~10$^4$~cm$^{-3}$) \citep{blac1987, gatl1987, dine1988}.  However, \cite{burt1990} found UV fluorescent H$_2$ emission from clumpy photodissociation regions with shock-like values for the 2$\rightarrow$1/1$\rightarrow$0 line ratio and $n_{\rm H}$~$\ge$~10$^{5}$~cm$^{-3}$, suggesting that the value of 0.1 for the ratio is not unique for shock heated gas.  In addition to these two processes, the response of H$_2$ gas irradiated by X-rays has been studied by several groups who conclude X-rays are capable of stimulating H$_2$ emission in a variety of astrophysical environments, including circumstellar disks \citep{gred1995,malo1996,tine1997}.  Due to the collisional nature of the X-ray ionization process for stimulating H$_2$ emission, this mechanism also has been predicted to populate the excited rovibrational states of the H$_2$ molecules such that the $v=2\rightarrow1~S(1)/v=1\rightarrow0~S(1)$ ratio also has a value of $\sim$~0.1.  Therefore, given the apparent degeneracy of the line ratio, differentiating between these three mechanisms will also require the knowledge of the density and kinematics of the emitting gas.

\cite{nomu2007} present a detailed disk model that incorporates irradiation of the disk by X-rays and UV photons associated with the central source and includes dust grain growth and setting towards the midplane.  Using this disk model, \cite{nomu2007} calculate the level populations of H$_2$ in the model disk for a range of dust grain sizes and predict values for the $v=2\rightarrow1~S(1)/v=1\rightarrow0~S(1)$ H$_2$ line ratio for three stimulation scenarios: a) X-ray and UV, b) only UV, and c) only X-ray.  The values \cite{nomu2007} calculate for the line ratios differ significantly from the nominal values mentioned above and depend on sizes of the dust grains (see Figure~17 of \cite{nomu2007}).  They also find that as the dust grains grow and the grain opacity decreases, UV pumping should become the dominant stimulation mechanism over X-ray ionization.  According to the findings of \cite{nomu2007}, an accurate measurement of the line ratio will determine the mechanism responsible for stimulating the H$_2$ emission as well as indicate the evolutionary status of the dust grain population.

The upper limits of the 2.2477~$\mu$m line observations mentioned in \S\ref{res} can be used to determine an upper limit on the 2$\rightarrow$1/1$\rightarrow$0 line ratio for CS~Cha and HD~97048.  Unfortunately, with upper limits for the ratio of 0.4 and 0.5 for CS~Cha and HD~97048, respectively, these observations lack the sensitivity in both cases to test the predictions of \cite{nomu2007} and do not rule out the `only UV' (scenario b) case.

Simultaneous observations of both lines made with a large aperture telescope and high resolution spectrometer (R~$\ge$~60,000) will be ideal for accurately measuring this line ratio.  The simultaneity of these observations are important if the H$_2$ emission is variable, as non-simultaneous observations would then likely produce inaccurate measurements of the line ratio.  Indeed, the H$_2$ emission is likely to be variable if the X-ray and UV fluxes, both known to be produced by highly variable phenomena in these young systems, are responsible for stimulating the H$_2$ molecules.

\subsection{Transitional Disks}
\label{trans}

In the infrared, the shapes of the spectral energy distributions (SEDs) of cTTS systems are described as `flat' due to emission from circumstellar dust grains in excess of the Rayleigh-Jeans tail of the blackbody curve observed for the stellar photospheres of main sequence stars.  However, infrared photometric surveys of many TTS in nearby star forming regions, similar to those discussed in \S\ref{intro}, also identified dips in the 1-10~$\mu$m regions of the otherwise flat SEDs associated with some of their targets \citep{stro1989,skru1990,gauv1992}.  These findings suggest that the small warm dust grains responsible for the short wavelength infrared excesses in most cTTS disks are not present within the disks of these systems, while the cooler dust grains responsible for the longer wavelength excess persist in these disks.  Therefore, the observed dips in the SEDs of TTS have been interpreted as evidence of holes or clearings in the innermost regions of these disks.  These holes may arise due to the coagulation of small dust grains into planetesimals or the presence of a massive planet that dynamically cleared the inner regions of the disks \citep{skru1990,naji2007a}.  Additionally, disk photoevaporation models, in which disk material is heated and `boiled' away by highly energetic X-ray and UV photons, also have been offered as a phenomena responsible for the clearing of the inner holes in these transitional disks \citep{clar2001,alex2006b,mcca2006}.

For a few such transitional disk sources with inner holes extending out to several tens of AUs, such as GM~Aur and CS~Cha, tiny amounts of small dust grains,  $\sim$~0.02 and 5~$\times$~10$^{-5}$ lunar masses, respectively, located within $\sim$1~AU, have been detected \citep{calv2005,espa2007}.  Somewhat surprisingly, several of these transitional disks possess well-known accretion signatures (i.e., $U-$band excess, strong H~{\scshape i} recombination lines, etc.) suggesting that gas must remain within the disks and the inner holes, despite the lack of detectable dust emission.  \cite{chia2007} show that the accretion activity in these sources may be driven by coronal X-rays activating the magneto-rotational instability in the gas residing at the dust disk rim and residing within the inner disk hole.

\subsubsection{H$_2$ Emission from Transitional Disks}
\label{h2intd}

All sources toward which we have searched for H$_2$ emission \citep{wein2000,bary2002,bary2003a} known to possess transitional disks have been found to produce detectable levels of quiescent H$_2$ emission at 2.1218~$\mu$m, including TW~Hya, LkCa~15, CS~Cha, and GM~Aur (\cite{wein2005}); Bary et al.\ 2008, in prep).  In each case, kinematic information from the emission lines, assuming a similar velocity dispersion as found for the other H$_2$ detections and for the unresolved feature measured for the disk of TW~Hya\footnote{The detection of H$_2$ emission from TW~Hya is reported as an unresolved detection in CSHELL spectra taken at the IRTF.  The FWHM of the spectral feature was incorrectly quoted as $<$~6~km~s$^{-1}$ or the 1$\sigma$ width.  The FWHM of the unresolved line is actually measured to have a value of $\sim$~13.5~km~s$^{-1}$, the instrumental resolution of the instrument.} \citep{wein2000}, places H$_2$ gas within the optically thin inner disk hole for each of these three sources.  \cite{rett2004} detected $v=1\rightarrow0$ rovibrational CO emission from the disk of TW~Hya, but did not detect the $v=2\rightarrow1$ CO band.  \cite{rett2004} suggest that the absence of `hot' CO gas indicates that the inner 0.5~AU of the disk has been cleared.  A rotational temperature of $\sim$~430~K leads these authors to conclude that the CO emission they detected in this system arises in warm gas occupying a very narrow annulus in the disk ranging between 0.5 and 1~AU.  More recently, \cite{saly2007} detected warm CO emission from the disks of TW~Hya and GM~Aur, further confirming the presence of molecular gas in the inner regions of these transitional disks.  Several of the other six sources detected in H$_2$ are classified as borderline c/wTTS or wTTS, including Sz~33 and DoAr~21 \citep{bary2002}, which also makes them viable transitional disk candidates.  Given that we estimate the innermost radius for the location of the gas producing the H$_2$ emission in these sources to be near $\sim$~1~AU, we conclude that these inner disk holes are not completely devoid of gas, in agreement with the detection and measurement of accretion activity in several sources possessing transitional disks.

We must consider whether this apparent correlation of H$_2$ emission with the presence of transitional disks could be the result of observational bias.  Specifically, is H$_2$ emission easier to detect toward transitional disk sources because their near-infrared continuum flux levels are lower and thus do not overwhelm the H$_2$ line emission?  If all cTTS generate some amount of quiescent H$_2$ emission, but sources with flat spectra and a resulting higher near-infrared continuum flux make detection of these narrow and weak spectral features more difficult, then the brightest $K$-band sources should be among the hardest toward which to detect quiescent H$_2$ emission.  HD~97048, Glass~Ib, and CV~Cha, all of which are detected in H$_2$ emission, however, are among the brightest $K$-band sources in our sample.  Therefore, we suggest that sources possessing transitional disks are preferentially detected in H$_2$ line emission because of physical conditions and phenomena intrinsic to these objects, rather than observational selection.

As discussed in \S\ref{ratio}, grain sizes have a notable effect on the cross-sections for UV and X-ray photons; consequently, as disks evolve and dust grains grow larger, the locations in a disk to which the photons penetrate and deposit their energies change \citep{nomu2007}.  As a result, the TTS that are sources of quiescent H$_2$ emission may be those that have disks that permit photons to penetrate to regions where the densities permit successful stimulation of H$_2$ emission.  These disks would preferentially be those with optically thin inner regions and optically thick outer regions, i.e., transitional disk sources.  For sources that are both X-ray and UV sources, a real dependence of the stimulation of quiescent H$_2$ emission on the size distribution of the dust grain population argues for UV fluorescence as the dominant stimulation mechanism, as UV cross-sections are much more dependent on the sizes of the dust grains.

\subsection{The H$_2$ Emitting Region}
\label{emr}

In this section, we address the question of the location of the emitting H$_2$ gas stimulated by X-rays and/or UV photons, in the context of a transitional disk.  We predict the regions of such disks in which the competing stimulation mechanisms will be most effective by finding areas of the disk bounded by optical depth surfaces associated with energetic photons.  We then compare these regions with those determined from the kinematics of the emission features.

\subsubsection{UV \& X-ray Optical Depth Surfaces}
\label{opds}

Previous spectroscopic data \citep{bary2003a} and those presented in this paper suggest that the emitting gas extends from $\ge$~1~AU out to intermediate disk radii of 10-30~AU, with the bulk of the emitting gas residing at intermediate disk radii.  We seek to determine if any agreement may exist between these empirically determined radii for the gas emission and those likely to be associated with emitting gas produced via the UV fluorescence and X-ray stimulation mechanisms.

Utilizing a standard TTS disk model \citep{glas1997}, H$_2$ cross-sections found in \cite{yan1998}, and ignoring the contribution of dust and other trace gas components, to first order, we estimate where the UV photons and X-rays are absorbed and are likely to stimulate H$_2$ molecules in an atypical TTS disk devoid of small dust grains by calculating a range of optical depth surfaces for X-rays and UV photons with different energies impinging upon a disk with varying angles of incidence.  The physical depth into the disk that each of these photons penetrates is determined by

\begin{equation}
\tau = \int_{r_{inner}}^{r_{\tau}} \kappa_\nu \left(\frac{{\rm r}}{{\rm r_{0}}}\right)^{-{\rm q}} {\rm dr} ,
\end{equation}

\noindent
where $\tau$ is the optical depth, r$_{inner}$ is the radius at which the photons enter the disk, r$_{\tau}$ is the penetration depth of the photon for the given optical depth, r$_0$~=~1~AU, q is the power-law index (here taken to have a value of 2.75), and $\kappa_\nu$ is the opacity of the H$_2$ for photons with frequency $\nu$.
The photons are assumed to enter the disk at 0.35~AU from the star, the innermost radius for which the model applies.  After calculating the penetration depths (r$_{\tau}$) for a range of energies and optical depths, plots of the optical depth surfaces were constructed (Figures 2 \& 3) to illustrate the most likely regions of the disks in which the photons would be absorbed.  The plots consist of density contours derived from the \cite{glas1997} disk model overlaid with the optical depth surfaces.  The thin black lines represent density contours derived from the disk model, while the bold black lines represent the optical depth surfaces.  In Figure 2, we outline the surface corresponding to the uppermost boundary of penetration by either UV photons or X-rays (i.e., the $\tau$~=~1.0 surface for a UV photon with energy 15.4 eV, corresponding to a wavelength of $\sim$~805~\AA).  This is the lowest energy for which the cross-section of an H$_2$ molecule can be determined using the models of \cite{yan1998}, and comes closest to matching the photon energies used in the UV fluorescence models of Black and van Dishoeck (1987).  
 
In Figure~3, two optical depth surfaces are plotted with the upper bold contour representing for $\tau$~=~0.2 for UV photons with the same energy as in Figure~2 and the second contour at lower disk elevation representing $\tau$~=~1.0 for X-ray photons with an energy of 1~keV.  Given the greater penetration depths associated with the X-rays, the $\tau$~=~1 optical depth surface acts as the lower boundary for the penetration depth of the high energy  photons.  As illustrated by the plot, some of the X-rays are capable of penetrating into regions of the disk in which the densities exceed the critical density, $\rho_c$ (where $\rho_c$~=~10$^{-16}$~g~cm$^{-3}$ or 3~$\times$~10$^7$~H$_2$~cm$^{-3}$; 4th model curve from the midplane) for the formation of the 2.1218~$\mu$m emission feature.  Although some TTS are observed to produce X-rays with energies as high as 10~keV, we do not include them on this plot since they would penetrate to even greater depths and densities far exceeding $\rho_c$.  We interpret the region lying between these two surfaces and above the critical density contour as location in the disk where the photons are most likely to deposit their energies and where the H$_2$ emission is most likely to originate.  Using the constraints placed upon the location of the emitting region by the spectral information, we deduce that the bulk of the emission comes from a region located 10-30~AU from the star and 5-15~AU above the disk midplane.

\begin{figure}[htbp]
\begin{center}
\includegraphics[angle=90,width=1.0\columnwidth]{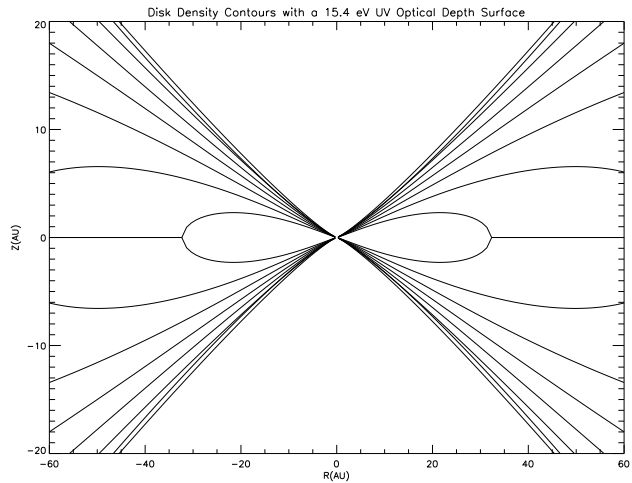}
\caption{An optical depth surface ($\tau$~=~1) for UV photons with an energy of 15.4 eV is plotted from 0.35 to $\sim$~45~AU in bold black contours.  The optical depth surface is overlaid atop density contours (thin contours) generated using the disk model of Glassgold et al.\ (2000).  The density contour nearest the disk midplane represents a density of $\rho$~=~10$^{-13}$~g~cm$^{-3}$.  With increasing elevation above the disk midplane, the values of the density contours decrease logarithmically, ending with $\rho$~=~10$^{-19}$~g~cm$^{-3}$.  The optical depth surface roughly corresponds to the distance into the disk that photons with a wavelength of 805~\AA\ could penetrate.}
\label{uvplot1}
\end{center}
\end{figure}

\begin{figure}[htbp]
\begin{center}
\includegraphics[angle=90,width=1.0\columnwidth]{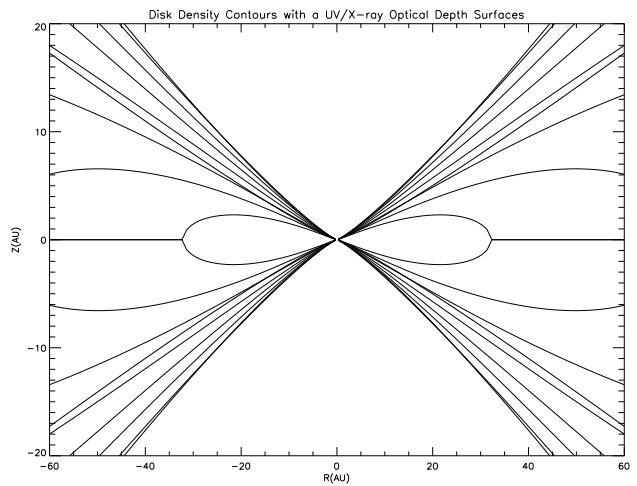}
\caption{The $\tau$~=~0.2 optical depth surface for 15.4~eV UV photons is plotted from 0.35 to $\sim$~45~AU as a bold black line atop the disk density contours.  The second bold black contour at lower disk elevation represents the $\tau$~=~1 optical depth surface for 1~keV X-rays.   The fourth curve from the midplane corresponds to 10$^{-16}$~g~cm$^{-3}$ or 3~$\times$~10$^7$~H$_2$~cm$^{-3}$ which is the critical density for the $v=1\rightarrow0~S(1)$ transition of H$_2$.  It appears that most of the X-rays will penetrate to densities $n_{\rm H_2}$~$>$~10$^{-16}$~g~cm$^{-3}$.
}
\label{littleopdplot}
\end{center}
\end{figure}
 






These calculations of optical depth surfaces demonstrate that the UV and X-ray photons most likely will deposit their energies in the upper levels of the disks at intermediate distances from the sources and confirm our earlier conclusions about the most likely location of the majority of the H$_2$ emission.

Having located the most likely region for the emitting H$_2$ gas, we next determine how much of the disk mass resides in the regions confined by the optical depth surfaces.  By integrating the density profile of the disk model provided in \cite{glas2000},

\begin{equation}
M_{disk} = \int_{r_{inner}}^{r_{outer}}\int_0^{2\pi}\int_{z_{lower}}^{z_{upper}} \rho_o \left(\frac{{\rm r}}{{\rm r_{0}}}\right)^{-{\rm q}} {e^{\frac{-{\rm z}^2}{2{\rm H}^2}}} {\rm r dr d\phi dz} .
\end{equation}

\noindent
where $r_{inner}$ and $r_{outer}$ and $z_{lower}$ and $z_{upper}$ are the limits of integration for the radial and vertical extents of the portion of the disk to be considered, $\rho_0$ is the density of the disk at a radius of 1~AU and is equal to $1.4\times10^{-9}$~g~cm$^{-3}$, and H is the vertical scale height parameter given by H$_0$(r/r$_0$)$^{1.25}$, where H$_0$~=~5~$\times$~10$^{11}$~cm and r$_0$~=~1~AU, we determine the mass of H$_2$ contained in the relevant regions outlined by the optical depth surfaces.  Using Equation~2 and setting the inner ($r_{inner}$) and outer ($r_{outer}$) radial limits to 10 and 30~AU, respectively, and approximating the vertical limits, $z_{lower}$ and $z_{upper}$, from the optical depth plots to be 5 and 10~AU, respectively, we find a mass of 2.8~$\times$~10$^{-5}$~M$_\odot$.  This mass is more than enough to account for the observed emission.  For comparison, only 5.8~$\times$~10$^{-9}$~M$_\odot$ lies in an upper region extending from 10~$\le$~z~$\le$~15~AU\footnote{Masses determined for the gas in the region of the midplane (0~AU) to 5~AU can be found in Table~3, but are not included in this discussion because the majority of this mass lies below the critical density contour, where no emission from the $v=1\rightarrow0~S(1)$ can be produced.}.  Since neither stimulation mechanism is 100\% efficient, a mass larger than that contained in the horizontal layer from 10 to 15~AU should be expected to reside in the region of the disk where the emission originates.  Therefore, the mass determined for the region 5 to 15~AU above the disk midplane suggests that the emitting gas likely extends downward, towards the midplane of the disk, with a minimum vertical height on the order of 5~AU at radii extending from a few to a few tens of AU for X-ray/UV stimulated H$_2$ emission in an accretionally-evolved disk associated with a TTS.

In our calculations of the optical surfaces for UV and X-ray photons, we have assumed that the gaseous component of the disk has been unaffected by the `disappearance' of the micron-sized dust particles.  If the gaseous component of the disk were to become more tenuous as the disk evolves, the resultant density structure would allow X-rays and UV photons to penetrate to even greater depths into the disk than suggested by the optical depth surfaces shown in Figures~2-6 and the critical density contour would move closer to the disk midplane and closer to the star.  In this case, which is similar to that expected for a transitional disk, the emission could arise from gas in regions of the disk located much closer to the midplane and closer to the star, and from gas at even larger disk radii.

However, in the case of an evacuated inner disk, but a dust-filled outer disk, the opacities will increase in regions of the disk still populated by dust grains.  Consequently, the shapes of the optical depth surfaces will change from those we have calculated.  We note that the extent of these inner holes is observed to vary from source to source and that these holes have observed sizes ranging from a few AU to $\sim$~40~AU, suggesting slightly different conditions for determining the location of the emitting gas for individual sources.

Although the optical depth surfaces calculated in this section provide insight into the location of the emitting gas from an accretionally evolved disk, a calculation of these same surfaces employing the irradiated disk models of \cite{nomu2007} and transitional disks, including a variety of cross-sections that incorporate the evolution of the size of dust grains will improve upon this investigation and make them applicable to a broad variety of accretionally-evolved circumstellar disks.

\begin{deluxetable}{ccc}
\tablewidth{0pt}
\tablecaption{Masses in H$_2$ Source Regions}
\tablehead{
			\colhead{$r_{inner}$$\rightarrow$$r_{outer}$}
		&	\colhead{$z_{lower}$$\rightarrow$$z_{upper}$}
		&	\colhead{H$_2$ Mass} \\
			\colhead{(AU)} 
		&	\colhead{(AU)}
		&	\colhead{(M$_\odot$)} \\
}
\startdata

10$\rightarrow$30                   &  		0$\rightarrow$5       &  6.4$\times$10$^{-3}$  \\
10$\rightarrow$30  		    &  		5$\rightarrow$10      &  2.8$\times$10$^{-5}$  \\
10$\rightarrow$30  		    &  		10$\rightarrow$15     &  5.8$\times$10$^{-9}$  \\

\enddata
\end{deluxetable}
\subsection{H$_2$ \& \rm [Ne {\scshape ii}]}

The results of several investigations \citep{herc2007,lahu2007,espa2007,pasc2006,bary2008} have found four sources possessing [Ne~{\scshape ii}] at 12.81~$\mu$m that also possess transitional disks.  Of the four transitional sources possessing [Ne~{\scshape ii}] emission, three (TW~Hya, CS~Cha, and GM~Aur) also possess quiescent H$_2$ emission, while the fourth (DM~Tau) has not yet been observed as part of any H$_2$ survey.  The strong overlap between the transitional disk sources possessing both [Ne~{\scshape ii}] and H$_2$ emission suggests a correlation exists between these two emission features and may be related to the nature of the transitional disks.  In fact, because of the high initial ionization potential of 20.6 eV for [Ne {\scshape ii}] emission \citep {glas2007}, this line probably traces the same high energy photons we theorize are responsible for stimulating the quiescent H$_2$ emission in the near-infrared.  Furthermore, the FWHM measured for both the [Ne~{\scshape ii}] and H$_2$ in TW~Hya suggest that the emission from these two features arises at similar radii in the circumstellar disks.

\section{Individual Sources}
\label{sources}

\subsection{CS Cha}
\label{cscha}
CS~Cha was classified by \cite{hart1993} as a cTTS, based on a measured H$\alpha$ EW of 65~\AA, though Appenzeller et al.\ (1983; 13.3~\AA) and Luhman (2004; 20~\AA)  found smaller values.  The observed variability of H$\alpha$ for CS~Cha was previously suspected by \cite{heni1973}.  \cite{gauv1992} and \cite{pers2000}, using IRAS and ISOCAM, respectively, found that CS~Cha has little or no near-infrared excess but substantial mid- to far-infrared excesses.  The magnitude and variability of the H$\alpha$ EW measurements for CS~Cha strongly suggest that this source is still actively accreting with a variable accretion rate --- in support of this conclusion, \cite{espa2007} estimated a mass accretion rate of 1.2~$\times$~10$^{-8}$~M$_\odot$~yr$^{-1}$ for CS~Cha from the $U$-band excess, while \cite{john2000} estimated it be 1.6~$\times$~10$^{-7}$~M$_\odot$~yr$^{-1}$ --- despite continuum measurements that the disk possesses an optically thin inner region.

Using {\it Spitzer} IRS spectra and MIPS photometry at 24 and 70~$\mu$m, \cite{espa2007} modeled the SED for CS~Cha and found a best-fit model that includes an inner hole with a radius of $\sim$~43~AU, with a small amount of hot dust grains located between 0.1 and 1.0~AU.  Based on the Espaillat disk model, the variable H$\alpha$ measurements, the $U$-band excess, the strong [Ne {\scshape ii}] emission, and our detection of H$_2$ emission from the CS~Cha disk, we suggest that CS~Cha is an X-ray bright, actively accreting TTS possessing a transitional disk with a large optically thin inner hole, containing little dust and a significant amount of gas.  The dust grain population is likely dominated by large grains, hence the low near-infrared excess, while the gas contributes substantially to the variable accretion rate.

\subsection{CV~Cha}
\label{lkha}

LH$\alpha$~332-21 is a binary cTTS \citep{reip1993}, composed of CV~Cha and CW~Cha, having a large separation of 11\ptsec7 \citep{ghez1997}.  We observed only CV~Cha, the primary component, and have detected H$_2$ emission from this source.  CV~Cha has an H$\alpha$ EW of 60.5~\AA\ \citep{gauv1992} and exhibits weak continuum veiling in high resolution optical spectra obtained with UVES on the VLT \citep{stem2003}.   Extended UV emission has been detected in IUE observations of this source by \cite{broo2001}.  With a spectral type of G8V \citep{rydg1980,taka2003} and a Log L$_{\rm x}$ of 30.1~ergs~s$^{-1}$ \citep{feig1993}, CV~Cha is one of most massive sources in our survey and one of the strongest X-ray emitters in the Chamaeleon~I star forming region.  This source is not known to possess a transitional disk and, due to a large H$\alpha$~EW and the presence of the aforementioned continuum veiling, appears to be an actively accreting source.  It appears that this source possesses the necessary attributes, in that it is a source of both UV and X-ray photons, to stimulate the observed quiescent H$_2$ emission.  

\subsection{Sz 33}
\label{sz33}

Sz~33 has been classified as both a wTTS \citep{feig1989} and a cTTS \citep{feig2004} with a value of 9.5~\AA\ for its H$\alpha$~EW, placing it amongst the other transitional objects, ``stradling'' the line between a cTTS and wTTS \citep{gauv1992}.  \cite{gome2002} present a low resolution (R~$\sim$~500) spectrum of Sz~33 that contains Pa$\beta$ and Br$\gamma$ emission features, indicating that this source is still actively accreting.  Given the similarity of its ambiguous classification as a cTTS and/or a wTTS to other sources with detectable H$_2$ emission in our survey that possess optically thin inner disk regions, we suggest that the disk associated with Sz~33 also may be a transitional disk.

\subsection{HD 97048}
\label{hd97048}

HD~97048 is classified as a B9.5V Herbig AeBe star, one of two Herbig AeBe stars that were observed as part of this survey (HD~97300 was the other).  Our detection of quiescent near-infrared H$_2$ emission and the detection of pure rotational H$_2$ emission by \cite{mart2007} from this source places it in the company of AB~Aur, another Herbig AeBe star, for which pure rotational H$_2$ emission has been detected by \cite{bitn2007}.  Infrared emission from this source is thought to originate from the surface layer of a passively heated disk.  Extended mid-infrared emission is dominated by a strong continuum produced by warm dust grains in the immediate vicinity of the star and solid state emission bands from silicates and carbonaceous materials \citep{meeu2001,natt2001,prus1994}.  The spectrum rises rapidly longward of 2.2~$\mu$m and peaks around 60~$\mu$m \citep{hill1992,boek2004} and is similar to the SEDs of TTS possessing transitional disks.  An extended disk ranging from 360 to 720~AU was observed with the Advanced Camera for Surveys on the {\it Hubble Space Telescope}.  Ca {\scshape ii}, Na {\scshape i}, and O {\scshape i} are all observed in absorption suggesting that the source is neither actively accreting nor generating a bipolar outflow.  Despite the absence of the accretion activity necessary for the production of UV photons in the low mass counterparts, copious amounts of UV radiation intrinsic to the SED of an early-type star as well as its detection as an X-ray source demonstrate the presence of the underlying mechanisms for high energy photon production we believe may be responsible for stimulating the H$_2$ emission.

\subsection{Glass Ia \& Ib}
\label{glassI}

Glass Ia and Ib form a close binary system with a separation of 2\ptsec4, or a minimum physical separation of $\sim$~380~AU.  The secondary component is classified as an Infrared Companion (IRC) based upon the 2.2~$\mu$m flux ratio (F$_{2.2}$(optical primary)/F$_{2.2}$(IR companion)) \citep{natt2000} of 0.32.  \cite{hart1993} and \cite{natt2000} independently have categorized Glass~Ia/Ib as a wTTS with a measured H$\alpha$~EW of 4~\AA.   \cite{lomm2007} did not detect 3.3~mm continuum emission from Glass I, suggesting that there are little to no mm-sized dust grains in the circumstellar and circumbinary environment.

The detection of H$_2$ emission from Glass I is quite unique amongst our other detections since it represents the only source for which we have detected spatially resolved emission.  We have determined that all of the emission whether centered on the primary, the secondary, or in the extended emission, is centered on the stellar rest velocity with a dispersion of only ~1.5~km~s$^{-1}$, very near the estimated measurement error.  Therefore, we find that the emitting H$_2$ gas in the system displays a common and systemic velocity. 

When near-infrared H$_2$ emission is found to be spatially extended in the vicinity of a young star, the kinematics of the glowing gas typically reveal that it is associated with shocks in an outflow.  The lack of dispersion in the central velocities of the extended H$_2$ emission features observed from the Glass Ia and Ib binary system, however, argue against the outflow scenario and indicate that this gas is associated with circumstellar and circumbinary disks in this binary system.
Gas orbiting a 2~M$_\odot$ binary with a radius greater than 1700~AU will not produce a detectable velocity shift at this resolution.  We do find that the FWHM of the emission features centered on both sources are slightly larger than those measured for the extended gas, with FWHM of 12.6 and 17.6~km~s$^{-1}$ found for the primary and the secondary, respectively, and 10.3 and 9.1~km~s$^{-1}$ measured for gas located 0\ptsec8 south the secondary and gas extending 0\ptsec45 to the south of the primary component, respectively.  The difference in the FWHM measured for the components of the binary suggest that we are detecting emission from disks with different inclinations and/or central sources with different masses.  For the extended gas, with its location likely to be in a circumbinary disk, little broadening of these lines will be contributed by the Keplerian motion.  Assuming the standard thermal excitation temperatures for H$_2$ of 2000~K, the FWHM of the line is expected to be $\sim$~7~km~s$^{-1}$, slightly smaller than the FWHM measured for the extended emission features.  The measured FWHM correspond to unreasonably high temperatures greater than 3000~K, perhaps, suggesting the presence of some turbulent broadening in the extended gas emission.

Although it is not currently possible to reconcile the broader than expected FWHM of the extended emission features with their origin in a circumbinary disk, it appears we can rule out the association of this emission with outflow activity.  Since we find a larger FWHM for the lines centered on the components of the binary, and the FWHM is similar to those detected for the other detections in this survey, we infer that these features are associated with the circumstellar disks of the individual components.

\subsection{X-rays vs. H$\alpha$ EWs}
\label{xrayha}

In Figure~7, we present a plot of Log~L$_{\rm x}$ versus H$\alpha$~EW for all published H$_2$ detections and non-detections to determine if any correlation exists between these two parameters and the presence or absence of detectable quiescent H$_2$ emission.  If H$_2$ emission is generated by X-rays produced in coronae and by accretion activity on the surfaces of these stars, we expect a clear correlation to exist between sources with detectable H$_2$ emission and their X-ray luminosities.  Figure~7, however, shows that sources with a variety of X-ray fluxes, including a few sources without detectable X-ray fluxes, are among those with observed quiescent H$_2$ emission.  Assuming that large intervening columns of circumstellar gas are not responsible for extinguishing the intrinsic X-ray luminosities of the apparently X-ray quiet sources, the lack of any correlation between H$_2$ detections and the presence and magnitude of a X-ray flux suggests that X-rays may not be responsible for stimulating the H$_2$ emission.  On the other hand, there appears to be some correlation between H$\alpha$ EW strengths and H$_2$ detections as all but a few sources in the region of the plot with H$\alpha$~EWs greater than 10~\AA\ (the rough boundary between cTTS and wTTS), produce detectable levels of H$_2$ emission.  This correlation between H$\alpha$ and H$_2$ emission links accretion activity with quiescent H$_2$ emission and are similar to the findings of \cite{carm2007}.  Since accretion activity in TTS is responsible for the production of the UV photons necessary to produce fluorescent H$_2$ emission, this correlation suggests that UV fluorescence is likely the dominant stimulation mechanism in some, if not all of the detections.


We now turn our attention to the transitional disk objects in Figure~7 denoted by the large asterisks overlaid upon the filled circles.  Each of these sources (LkCa~15, CS~Cha, and TW~Hya) has been shown to possess an inner disk hole that is either devoid of dust grains or possesses an evolved population of large dust grains.  In addition to these three transitional disk sources, HD~97048 has a similar infrared SED, as noted in \S\ref{hd97048}, suggesting that it too possesses a large inner disk hole.  If we consider that the other wTTS, Sz~33 and DoAr~21, are likely transitional disk objects, and, for the moment, ignore the binary sources GG~Tau and Glass~I due to their complexity, only three of the H$_2$ detections (CV~Cha, ECHA~J0843.3-7905, and LkH$\alpha$~264) now appear as cTTS with non-transitional disks.  However, the observations required to construct infrared SEDs do not currently exist in the literature.  Future observations will be required to further test this correlation.  Based upon the correlation we have found between transitional objects and quiescent H$_2$ emission, we conclude that objects detected in quiescent H$_2$ emission and designated as cTTS also may harbor transitional disks and that stimulation mechanisms responsible for producing the quiescent H$_2$ emission are likely related to both accretion activity and the evolutionary status of dust grains within their inner disks.  

In addition to the possible impact of the evolutionary status of dust grains, the presence of UV and/or X-ray photons, and accretion activity on the production of quiescent H$_2$ emission, variability may also play a role in the detectability of these H$_2$ emission signatures.  As in the case of most optical and infrared emission lines observed from TTS, the H$_2$ spectral features may fluctuate on similar timescales.  We expect this variability to have a negative impact on our detection rate and may explain why some sources that share favorable characteristics with sources that are detections do not appear to produce detectable levels of H$_2$ emission.  Therefore, variability may explain the lack of H$_2$ detections for some of the borderline cTTS seen in Figure~7.  In order to determine if this is indeed the case, multi-epoch observations of some of the detections will be required to measure the magnitudes and timescales for any variability that may be associated with these sources.  Simultaneous observations of the 2.1218~$\mu$m line as well as other accretion indicators also would be helpful in linking the emission with UV activity associated with accretion hot spots on the stars.

\begin{figure}[htbp]
\begin{center}
\includegraphics[width=1.0\columnwidth]{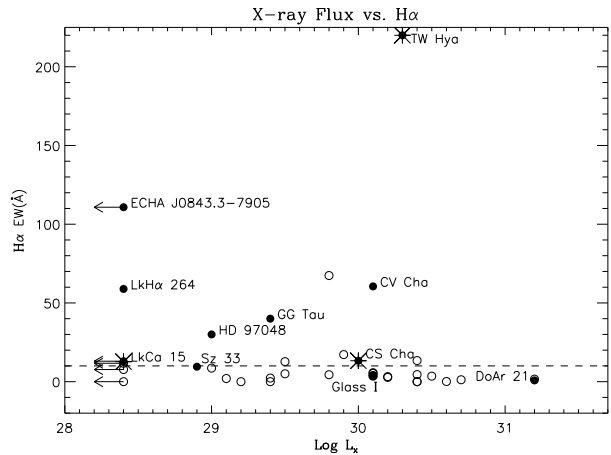}
\caption{Log L$_{\rm x}$ versus H$\alpha$ EW values for all sources with published observations of H$_2$ emission.   The detections are denoted with filled circles and non-detections with open circles.  Objects with known transitional disks are denoted by an asterisk.  The horizontal dashed line designates the canonical dividing line at 10~\AA\ separating cTTS (above) from wTTS (below).  Arrows pointing to the left designate the sources which were non-detections in X-ray surveys.  Detections are labeled with source names.}
\label{halphxray}
\end{center}
\end{figure}

\section{Conclusions}

As part of a high resolution near-infrared spectroscopic survey for H$_2$ emission from TTS in nearby star forming regions, we have made new detections of emission features centered on the 2.1218~$\mu$m $v=1\rightarrow0~S(1)$ transition of H$_2$ in the spectra of two cTTS (CS~Cha and CV~Cha), two wTTS (Glass Ia and Sz~33), one IRC (Glass Ib), and one Herbig AeBe star (HD~97048) located in the Chamaeleon~I dark cloud.  The detection of H$_2$ we report toward HD~97048 is the first detection of near-infrared emission from quiescent H$_2$ toward a Herbig AeBe star.  In every case, the central velocity of the emitting gas is found to be coincident with the systemic velocity of the source.  Assuming a circumstellar disk origin for these emission lines, an assumption well-supported by the central velocities of these lines falling at the rest velocities of the respective stars and by the narrow line widths, we used the HWZI and the FWHM velocities of these Gaussian-shaped emission features to estimate the innermost radii to which the gas extends and the radii at which the bulk of the emitting gas resides .  We find that the innermost radii for all of the stars is on the order of $\sim$~1~AU.  We further find that the radii about which the bulk of the emission is centered is on the order of $\sim$~10~AU.  For the binary source (Glass I), H$_2$ emission extends between both components and to the south of the primary.  Although the extended emission suggests that the gas could be entrained in an outflow, the lack of a detectable velocity gradient along the would-be axis of the outflow makes this an unlikely scenario.  However, a circumbinary disk origin for the extended emission is also difficult to reconcile with the large FWHM of the emission lines measured for the extended emission.

The initial goals for this survey was to determine if wTTS lacking detectable emission from standard disk tracers still harbor protoplanetary disks capable of or in the process of forming planetary systems.  Our survey revealed two such stars, making a total of three, including DoAr~21 discovered by us in a previous survey.

The range of H$_2$ line fluxes (6.9$\times$10$^{-16}$~$\le$~F$_{2.12}$~$\le$~8.6$\times$10$^{-15}$~ergs~cm$^{-2}$~s$^{-1}$) and masses (10$^{-10}$~$\le$~M$_{H_2}$~$\le$~10$^{-11}$~M$_\odot$) measured for the Chamaeleon~I sources with detected H$_2$ line emission were found to agree with all previous detections of quiescent H$_2$ emission.  These trace amounts of gas do not directly inform us about the total disk masses harbored by these systems, though almost certainly the mass of H$_2$ detected directly through line emission at 2.1218~$\mu$m represents only a small fraction of the total mass in these disks.  Our disk models suggest that the total disk masses could be 10$^2$ to 10$^7$ times greater than the mass measured directly in this single emission line.

Assuming that shock excitation is not the most likely mechanism for stimulating narrow line emission from H$_2$ gas which is not Doppler shifted from the systemic velocity of the source, we suggest that UV fluorescence and/or X-rays are the most likely stimulation mechanisms responsible for the observed H$_2$ emission.  We calculated optical depth surfaces for UV photons and X-rays penetrating a disk composed entirely of H$_2$ to determine where in a core-accretionally evolved disk these photons deposit their energy and possibly generate H$_2$ emission.  These models confirm that the radiation is most likely to be absorbed by gas in the upper atmospheres of these disks at intermediate radii and show that these stimulation mechanisms require significantly dense and massive columns of gas to produce the observed emission features at radii coincident with the those determined from the kinematic information.  Many improvements may be made to these calculations by including opacities associated with large dust grains, other gaseous species, or a disk model that better approximates an irradiated transitional disk.  However, the results from these calculations suggest that a gas-rich inner disk is necessary to produce quiescent H$_2$ emission via UV photons and X-rays, and that the stimulated gas resides at high disk altitudes and intermediate disk radii.

All four sources in our survey to-date known to possess transitional disks with optically thin inner disk holes are also H$_2$ detections.  These detections place gas within the inner disk holes similar to the resolved and unresolved detections of [Ne {\scshape ii}] emission lines from the sources.  The presence of gas in the inner disk holes confirms the presence of gas within these gaps necessary to explain the accretion activity observed for these sources.  The correlation between sources possessing quiescent H$_2$ emission and [Ne {\scshape ii}] suggests a shared stimulation mechanism for both of these lines, a relationship to the photoevaporation process, and a distinct stage in the evolution of these circumstellar disks.  We also find that most sources in our survey with H$\alpha$~EWs sufficient to classify them as cTTS have detectable H$_2$ emission.  The only exception is WW~Cha, which may also be a detection, though we have cautiously classified it as a `possible' detection, sine the H$_2$ emission towards this source, if real, is only at the 2.4$\sigma$ level.  The apparent relationship between H$\alpha$ emission and accretion activity suggests that the quiescent H$_2$ emission may likewise be correlated with accretion in agreement with the findings of \cite{carm2007}.  This is not entirely surprising given that the UV photons necessary for stimulating H$_2$ emission are likely generated by accretion hot spots on the surfaces of the stars.  In fact, this is strong evidence that UV fluorescence undoubtedly contributes to the observed H$_2$ emission.  Given the dependence of the UV penetration depths on the sizes of the dust grains in the disk, illustrated by \cite{nomu2007}, and the overlap between H$_2$ detections and sources possessing transitional disks, we suggest that the evolutionary stage of the dust grain population in the inner regions of these disks is also correlated with the production of quiescent H$_2$ emission.  Based upon these findings, we propose that Sz~33 and DoAr~21, both H$_2$ sources that are borderline c/wTTS, likely harbor transitional disks.

\acknowledgments

The authors would like to thank the staff members of Gemini South who were involved in collecting the data for this survey, especially Bernadette Rodgers, Phil Puxley, and Steve Ridgeway.  The authors would also like to thank the anonymous referee for useful comments that improved the manuscript considerably.  JSB acknowledges support from a NSF Astronomy and Astrophysics Postdoctoral Fellowship grant AST-0507310.  JSB, DAW, JHK, and SJS acknowledge support from NASA Origins of Solar Systems (NAG5-13104) and from NSF AAG (AST-0307416).  



\end{document}